# Deleuze's *Postscript on the Societies of Control*
## Updated for Big Data and Predictive Analytics
*James Brusseau*

**Abstract**: In 1990, Gilles Deleuze published *Postscript on the Societies of Control*, an introduction to the potentially suffocating reality of the nascent control society. This thirty-year update details how Deleuze's conception has developed from a broad speculative vision into specific economic mechanisms clustering around personal information, big data, predictive analytics, and marketing. The central claim is that today's advancing control society coerces without prohibitions, and through incentives that are not grim but enjoyable, even euphoric because they compel individuals to obey their own personal information. The article concludes by delineating two strategies for living that are as unexplored as control society itself because they are revealed and then enabled by the particular method of oppression that is control.

In 1990, Gilles Deleuze published *Postscript on the Societies of Control*, a milestone in philosophy's application to culture, economics, and advancing technology. The essay is short, speculative, and divided into three sections. The first describes the control society as nascent, and then delineates it in historical terms by contrasting it against the preceding disciplinary society. Section two outlines the control society logic as a set of premises, behaviours and concepts. The last section initiates a programme for living in a society of control. This essay updates Deleuze's *Postscript* with three aims. First, to show how the mechanisms of social control have developed over the last thirty years. Today, 'control' is no longer an abstract concept in political philosophy; it is localisable as specific technologies functioning where personal information is gathered into contemporary data commerce. Second, to develop the core element of Deleuze's definition of control: social organisation without spatial divisions and explicit prohibitions. The most pressing question Deleuze asks is, *How can there be control if nothing is forbidden?* The answer is predictive analytics: data-driven marketing and social media strategies that regulate through incentives. Consumers are not contained, blocked, or forced, but they are lured and directed by marketing appeals exploiting personal information. The third aim of this essay is practical: it is to explore how technology users can respond to the control society.

Since Deleuze published his short essay, secondary literature has developed in multiple directions. The most prominent contributions involve political and cultural realities of control (Wiley and Wise 2019; Hardt and Negri 2017; Konik 2015). There has also been a movement to connect the essay with the neglected philosopher Gilbert Simondon, and questions about personal identity (Hui 2015). Issues in biopower have arisen (Cohen 2018; Nail 2016). Debates in urban planning have also revealed interesting connections (Hagmann 2017; Galič et al. 2017; Krivy 2018). Though work has been done in the area of control and the economics of wealth inequality (Barthold et al. 2018) this article is, to the author's knowledge, the first to consider Deleuze's ideas within the economic context of marketing, information processing, and predictive analytics.

Finally, the triptychal structure of Deleuze's original composition is maintained here. Each of the three parts begins with a brief review of the original essay section, and then recontextualises the ideas within the big data reality.

## Historical

Deleuze began by drawing historical distinctions between disciplinary and control societies. Corresponding with the Western industrial societies of the eighteenth through twentieth centuries (Foucault 1977), the disciplinary regime organises rigidly, by sheltering, enclosing, and confining people in demarcated spaces dedicated to specific activities. Control, which is ascendant in the West, works fluidly: instead of walls and boundaries, social organisation is accomplished by mapping and channelling our movements (Haggerty and Ericson 2000).

Penitentiaries contrasted against the electronic ankle bracelets of home arrest exemplify the discipline and control distinction. Being enclosed by penitentiary walls identifies you as a convict and compels activities by location: when sent to the cafeteria, you eat, when released to the yard, you exercise. This is discipline in its starkest form, place *is* social imposition. The ankle bracelet organises differently. Those sentenced to wearing the device as home arrest may eat in whichever room, at whatever hour: the walls surrounding this kind of inmate make few demands. But, the weakness is offset when movements are remotely tracked, when the data is collected and analysed to produce a sense of the bracelet's wearer. The processing configures more than spatial coordinates and hourly schedules, it is also personal habits, biological needs, human desires. Control starts here, as the organisation and analysis of movements and patterns, and as the data increases, concrete walls become redundant: at least some convicts are no longer worth enclosing when wardens know where they are, and can predict where they're headed.

Any comparison between two historical periods suggests this question: Which is preferable? There may be an answer, but for Deleuze discipline versus control does not create a hierarchy so much as distinct contexts for a continuing tension between social oppression and personal freedom (Dreyfus and Rabinow 1983). On the disciplinary side, the fundamental struggle is about *how space is used*: it is whether a long iron pipe running underneath the prison is a sewage channel, or an escape route. Correspondingly, when the tension circulates through the ankle bracelet, the conflict applies to *patterns of information and their disruption*. Suffering constraint – and being free – involves struggling to control access to data about where I am, when, and where I may be expected to go.

### Thirty-year Update

Thirty years later there are updates to Deleuze's essay on the levels of technology, mood, psychology, and strategy. Each one demonstrates how Deleuze's vision of social organisation without walls and constraints has developed rapidly.

Starting with technology, Deleuze speculated: 'A mechanism giving the position of any element within an open environment is not necessarily one of science fiction.' (Deleuze 1992: 7) Regardless of how fictional it sounded then, today these mechanisms are as common as mobile phones constantly signalling the nearest transmission towers. The same pinging ensuring calls and text messages without interruption also constantly reveals the location of the individuals corresponding with the numbers (Taylor 2015; Ghose 2017: 2–17).

More significant than the emergence of geolocation into quotidian experience is *how many* locaters there are. When Deleuze started, he foresaw a future city where natives carried e-cards to swipe in the morning when leaving home, and then when entering the office, and when checking out at lunch, when stopping at a restaurant in the evening. Today, that single card has converted into multileveled gathering by multiplying gatherers: contemporary technological reality is wildcatting data collection. There are toll tags in cars, and license-plate cameras at intersections (McIntyre et al. 2015: 13–18). Foursquare asks us to check in at different places we visit. Yelp tells us what's nearby and therefore what we are near. If we search Google Maps, then knowledge about our destination is ready for dissemination even before we leave. While that locating is occurring, other opportunistic gatherers contribute distinct data sets (Fan and Gordon 2014: 74–81). LinkedIn tracks employment histories, Amazon compiles reading habits, Facebook scrapes vacation stories, Tinder chronicles romantic tastes, retailers catalogue shopping histories (Willems et al. 2017: 228–242).

Next, data brokers with less than familiar names like Acxiom and LiveRamp acquire the information and begin cross-referencing. By finding a common phone number they unite someone who uses Uber regularly, with that same someone who spends a lot of time on Tinder. Similar connections join usernames, overlapping addresses, and resembling faces. As the information links the various aspects of a life (professional, recreational, romantic), facets of identity initially dispersed across multiple platforms stack into a single silo (Glasgow 2018: 25). So, it is not just the amount of observational data that outstrips Deleuze's original vision, it is also the source variety.

The contemporary business of intersecting names across data sets is titled *identity resolution* by one industry leader (Kobayashi and Talburt 2014: 349–354), and it aims for endlessly robust personal profiles. There is no end to the resolving because the identifying information that data brokers compile gets sold right back into the marketplace that first yielded the knowledge (Zuboff 2019: 233–255). As voracious social media platforms (LinkedIn, Tinder) and online vendors (Amazon, Wayfair) buy statistics revealing their users' commutes, necessities, interests, habits, weaknesses, and aspirations, they refine their services to deepen engagement and increase sales. And those transactions create still more data on the level of the platforms which once again can be sold up to the data brokers, further compiled, and then churned back into the economic machine in an accelerating cycle (Aloysius et al. 2016: 1–27).

So, after Deleuze, the technology of control has reformed. Data gathering is no longer a single e-card, but multiple and multiplying mechanisms. Data compiling is no longer centred on one gatherer, but dispersed across corporations and enterprises. Data recording is now subordinated to synthesising information from multiple sources. And data's end result is not accumulation, but redistribution at an accelerating pace back into the profitable networks that first produced the information.

Turning from technology to mood, there was an anguish in the original essay. It is up to us, Deleuze warned, 'to discover *what we're being made to serve*' when our lives are subjected to the control society (Deleuze 1992: 7). The essay also cites Kafka, who is inseparable from his dejected imagination, as well as William Burroughs who, besides being an author, was a heroin addict, murderer, and object of constant police surveillance (García-Robles 1995). While the tone of writings by many authors in social and political philosophy can be tinged with apprehension, it is rarely so grimly determined as in Deleuze. '*There is no need*', he intoned, '*to fear or hope, but only to look for new weapons*' (Deleuze 1992: 4).

Today, the question is: *New weapons to fight whom?* The idea of agony and struggle *against* forces of social organisation and the powers-that-be has not aged well. The providers of social order are no longer grim oppressors, they are more likely cheerful marketers and colourful social media platforms. They are not bitter administrators and passive-aggressive bureaucrats issuing e-cards to all citizens and requiring their use as demanded by the jack-booted powers of a monopolising corporation or political state, as may be developing in China today (Kostka 2019). Instead, they are energetic professionals, sharply dressed, backed by spreadsheets, supported by software coders, and dedicated to accommodating our wants. They don't stand *against* the people, but in *favour* of consumers and users. If today's forces of social order are against anything, it is other marketers, other retailers, other content suppliers and social media platforms. What Deleuze perceived as a world of *oppression*, has become a reality of *competition*.

Just as strength and size (Sears, Kodak, bureaucracy, paper) have been replaced by speed and agility (Amazon, Instagram, algorithms, data) at the heart of economic success over the last thirty years, so too the social anxiety over centralised powers grinding against the people's resistance has been replaced by the excitement of marketplace innovation and the constant introduction of new digital applications. Users pass hours thumbing through potential romances on Tinder, with algorithms meticulously controlling to ensure that suiters are having fun, staying on the site, occasionally matching, and always providing still more information about their hopes and urges that can be sold on the downstream markets. The same mechanisms function on YouTube where viewers receive constant suggestions extending organically from what they are already watching. Facebook constantly tweaks and reprograms to keep people excited about the next round of selfie-illustrated gossip. If there is an underlying theme running through all this content provisioning, data collecting, and platform reengineering, it is enjoyment.

The result is that the brooding and surreptitious plotting that occasionally felt natural to post-1968 French political philosophy (Descombes 1980) has nearly evaporated. There is no centralised power to plot against, and control societies are not fearsome and inscrutable. 'Entertaining,' is more accurate. Of course, there is a threat inherent in the mass trivialisation of existence. The addictions of the algorithms dehumanise in their own ways, just as do the sweatshops of industrialisation. Regardless, it remains true that the machinations of contemporary control society operate on an emotional level that is more festive than ominous (Noyes 2019; Syvertsen and Gunn 2019).

Switching from mood to psychology, the updating of Deleuze's essay begins with a devious twist. For the future city where he imagined an e-card swipe granting entry to each location, Deleuze proposed, 'The card could just as easily *be rejected on a given day, or between certain hours*' (Deleuze 1992: 7). The scenario sounds like a futuristic version of Kafka's *Trial*, with an increasingly unhinged Josef K. obsessively returning to doors irregularly closed to him by a malevolent overseer.

Contrastingly, the reality today is that those professionals who track consumers and trade in the marketplace of resolved identities do not want to *observe and torment*, they want to *analyse and sell*. They are not twisted bureaucrats at Gmail pawing through inboxes and occasionally switching users' passwords to block account access, instead, they are strategists eager to discern patterns indicating imminent purchases. Because today's watchers want profit, not personal intimacy, they are interested in profiles, not specific people, and their attention is drawn to opportunities, not embarrassing information. If they are interested in vulnerability at all, then they mean vulnerable to *appeals*, like the new mother's enveloping concern for the health of her newborn, or the teenage boy's fixation on the size of his biceps. What comes next is not psychological persecution; instead, it is online coupons discounting a baby car seat, it is a protein drink free sample (Završnik and Levičnik 2015).

Stated differently, it is possible to read Deleuze's original essay and feel as though watchers may be focused on *me*. Today, though, with the tools of predictive analytics largely gathered in the hands of profit-seekers, the feeling that I as a human being am being observed (as opposed to analysed for economic opportunities) seems more like vanity than reality. There is also a numbers factor. With so many people generating so much data traffic – whether it is geolocating coordinates or intimate photographs – the anxiety felt by any one person that their personal data might escape the information deluge with sufficient magnetism to arrest the voyeuristic attention of some anonymous other, seems like a delusion of self-importance. Of course, people are watched today – as they always have been – by police and governments (Hu 2016), but that is not how *control* is working. Control is about data analysis, prediction, and economic opportunities, not voyeurism.

Pivoting from the watchers to the watched, the control update is equally transformative. Studies suggest that today's reaction to personal information gathering – especially among those most exposed, the social media youth – is neither anxiety nor angry paranoia, but something closer to the opposite. They either do not notice, or do not care about identifying data ogled by corporate surveillance, especially when a benefit is attached, like special discounts, or insider information about upcoming product offerings (Media Insight Project 2015).

If submission to the data inspecting of digital platforms yields rewards, it is even possible to imagine an aspiration or devotion to control. Why not actively reveal your browsing history, location data, and credit card purchases if that means your interests will be discerned and wishes efficiently met with packaged goods? Why not present your desires and fears for conversion into numerical maps for algorithmic processing in exchange for entertaining video suggestions from Netflix, attractive job opportunities from LinkedIn? If all medical records are compiled and cross referenced with eating habits, exercise history, and hereditary factors, then health outcomes improve, dramatically (Raghupathi and Raghupathi 2014). If dating services that once made do with questionnaires (Do you smoke? Drink? Blondes or brunettes?) could access terabytes of data involving romantic pasts (all the text messages, seduction patterns, weekend escapes, breakup episodes), then couldn't they discern something approaching guaranteed love?

None of these temptations imply that control has become less manipulative, only that control now is distinct psychologically from its origin: what started as obsession and dread has transformed into manic gratification.

Let us switch the update from psychology to strategy. Deleuze began envisioning control as movements tracked, and occasionally blocked by the maddeningly irregular locking of doors and passageways. However, in an immaculate control environment there are *no* impediments, not a single vestige of disciplinary enclosure. Pure control is limitless, obstacle-free. The ideal translates into a requirement to manage society's elements without blockages and prohibitions: a plan for getting out in front and *directing*. Perfected control does not only follow along, chart locations, and decipher patterns, it also leads.

According to *Tap: Unlocking the Mobile Economy*, today's advertising strategists mix big data and algorithms to:

> send a coupon to a potential customer before she even leaves
> for a shopping trip she didn't even know she was going to
> take. (Ghose 2017:1)

It could be that the woman was going to take the shopping trip regardless, but her record of habits and urges allowed analysis to predict the trip even before she did. Then, armed with the knowledge, a marketing tool was deployed to control what she would purchase: an enticing coupon appears on her Facebook feed. Or, still more proactively, it could be that her gathered personal information yielded the conclusion that she would be *vulnerable to the suggestion* of the trip, so the targeted offer arrives, and causes both the shopping excursion, and the specific purchase.

Either way, the result is a broader sense of how the strategy of control has advanced. First, as Deleuze described, knowing *where people are* (tracking them via their mobile phone as opposed to enclosing them somewhere) has converted into a form of social organisation. Second, tracking movements translates rapidly into *predicting* movements: as chartings overlay, patterns emerge (every Tuesday morning her credit card reveals a Starbucks purchase, shortly thereafter a sizable Whole Foods charge, later a toll camera snaps a licence plate registered in her name). The point is that for social regulation, the need for confinement diminishes as certainty increases about where people will be. Third, and most significantly, tracking and predicting give way to *targeting* (Walters and Bekker 2017). Targeting means offering incentives to specific individuals or delimited groups that have been tailored to appeal to their unique tastes, anxieties, and yearnings. When it works, people who were tracked and predicted become *directed*. What consumers and users are doing, why, and where they are headed later, all that becomes subject to control.

At this stage, control as an effective tool for social organisation may surpass discipline as a coercive power. Where discipline manifested as towering walls and locking doors, control is the strategy of leading and luring in a particular direction, and then in another. It is channelling and directing individuals without walls by triggering narrowly tailored desires and distributing opportunities strategically. This is LinkedIn displaying your profile in some cities and to some human resources managers, while obscuring it on other searches. Meanwhile, some curated restaurants are highlighted on your phone by OpenTable, while others get buried underneath uncommon search terms. Tinder grinds the patterns of attraction detected from your past swipes to spark romance, but it cannot work *too* well because that pushes users beyond the platform's control (assuming those in love delete their accounts). Everything is about enticing individual users, but only so far.

No matter the data platform involved, it folds neatly into a reality organised first by palpable incentives and simultaneously – though less tangibly – by the invisible force of opportunities foreclosed. No one is angry about the job they did not get because they were preoccupied by other opportunities and so never learned of the opening. No one's heart is broken by the person they never met because they were distracted by someone else. It is just important to note that these absences are not *denials*, they do not block like the walls of a prison. Instead, they are only the subjects of diverted attention. The strategic requirement for control today is *no explicit prohibitions*, no blocked possibilities, no forbidden ways. There are only opportunities and temptations. Some lead you towards, others lead you away.

All are nearly impossible to resist or escape. What can be resisted when nothing is being forbidden or denied? Where can escape lead when every direction converts into a trajectory of incentives?

## Logic

In 1990, Deleuze drew numerous conceptual distinctions between discipline and control societies.

Discipline operates through *discrete and separate spaces*, which means that physical thresholds impose behaviour shifts: no eating in the living room, no screens in the bedroom. Control associates with *geometric and continuous lines*, it is numerical and summative. Curves trace movements, mark times, and create patterns. The circulations, for example, of the e-card user Deleuze imagined could be mapped as the daily flow from home to work and back.

Discipline coheres with rigid *moulds*; one wage for the workers on the factory floor, another for the patrolling foremen, a third for executives sealed in offices. Control *modulates*: cash bonuses rise as workers speed more units down the assembly line. Managers receive stock options, with value fluctuating to reflect the company's success.

In a disciplinary society, we are always starting over again. Life is a constant reset as we convert from the bedroom, to the kitchen, to the workspace. Appropriate shoes are donned for work, later jeans because it is leisure time, then pyjamas for sleep. In control, things never begin or end, we are always in the middle. Deleuze remembers the ambiguous start and constant postponements of Kafka's *Trial*, and mentions the interminability of continuing education.

In a disciplinary society, we constantly reach for our identification. It allows us to drive our cars, get into bars, confirm the credit card as ours at the end of the night. Control society refers to a password. This combination of letters and numbers triggers the debit card, connects to the Minitel, the email.

The broad social cleavage for a disciplinary society contrasts the individual against the mass. Paradigmatically, the capitalist opposes the workers. In control society, the fundamental divide contrasts the individual against what Deleuze calls the 'dividual' (Deleuze 1992: 7). This dividual is the *me* resulting from summing all the data sets to which I belong. In my case, that includes male, 40–60, married, consumer of Coca-Cola, ex-smoker. The list goes on, but if followed to the end, the compiled *dividual* is the person writing this sentence.

Production anchors the economy of the disciplinary society: going to work in the morning means contributing to the fabrication of goods and the offering of services. The impetus is consumer *need*. Advertising – what Deleuze calls the *joys of marketing* (Deleuze 1992: 7) –drives the control economy. People go to work to sell. The impetus is consumer *desire*, which may be found and leveraged, or created and exploited.

### Thirty-year Update

Updates to the logic of control begin with time. In disciplinary societies, it was understood as segmented. There is a coherence here with cigarettes and alcohol: a smoke before entering the office in the morning, a drink after work; these chemical breaks facilitate a stop-and-start reality. As framing tools, the nicotine's quick rush and rapid dissipation, the drink's alcoholic wave across the consciousness, they viscerally break apart temporal stretches (Davies 2015). Today is different: fewer cigarettes and a different kind of time. Control-temporality never begins or ends because everything is always in process. Work emails get read in bed at 3 am. Leisure Friday became leisure everyday years ago, which does not only imply that work is becoming more comfortable, but also that the difference between work and leisure is becoming less definable. In the United States, that explains why leading technological organisations including Google and Apple are trading office buildings for lifestyle campuses (Morgan 2017). And now that fewer people in many advanced societies get married straight out of their school years, so too work life and romantic life entwine: where are you going to meet someone if not in the office? Unless, of course, it is a lover from years past who breaks into your Facebook feed. In the time of control, no relationship ever really ends.

Moving from time to personal identification, the transition from the card to the password that Deleuze noted has continued evolving towards facial recognition. In a sense, the face is the perfect password: impossible for the possessor to forget, and it does not matter whether others discover it. More, it is a culmination. Going back to a physical ID, like a driver's licence or a passport, the logic was the picture on the card corresponding with a name and description, and together they connected with the holder. There is an identification triangle. A password obviates the need to connect the identification with the holder because the password directly invokes the subject: it is no longer a question of visual resemblance, but of unique knowledge. So, three points reduce to two. Finally, with facial recognition, the identifier *is* the holder. The password *is* the person. There is no longer any distinction at all, I am my own identification, my own password. The convenience gain is significant, but the cost is the increased difficulty of *escaping* identification. In today's camera-laden world, we are revealing our password everywhere we go.

Turning to the subject of the social divisions Deleuze originally outlined between individual/mass on the side of the disciplinary society, and individual/dividual in the reality of control, that contrast has sharpened over time. Within the disciplinary, the weightiest oppressions emanated from the superhuman: the threat was dissolution of individuality into a larger *we*. Just as the factory dissolves individuals into a workers' union or an assembly line crew, so too in prisons, or schools, identifiable members fade into a collective in turn defined by the surrounding walls. Convicts melt into a cellblock, a student is a name on a class list. In both prisons and schools, the inmates/students are gathered and marched to their distinct places for defined activities: the cafeteria, the gym, the library. At the extreme, the spatial overlaps leave prisoners and students indistinguishable as their patterns of confinement replicate. This is what Deleuze referred to in the essay with his memorable if enigmatic citation from Rossellini's *Europa '51*, where the heroine views labourers (who may as well have been students) and exclaims, 'I thought I was seeing convicts!' (Deleuze 1992: 3).

The danger of control, by contrast, emanates from the *subhuman*. What threatens our ability to assemble a personal sense of who we are is the isolable *parts* of ourselves. In control society, *I* do not want a four-wheel drive Land Rover, instead, the fifty-year-old male, married, with incompletely attained professional aspirations wants that powerful vehicle. And, it is not *me* who wants to visit the Whitney Museum on Saturday, it is the person whose name appears on three lists: Land Rover owners, Manhattan residents, Amazon Alexa users. The Whitney marketing

department – with the help of big data operators like LiveRamp – has found that when those three slices of *dividuality* come together, the offering of a discounted membership likely receives an affirmative response. Of course the specific elements of dividuality and their corresponding marketplace opportunities are in constant flux, but what matters is this persistent reality: the gears of social control are not collectives and they are not individuals, they are narrow aspects of who we are (Lehn 2016; Schroeder 2016).

Where Deleuze uses the term *dividual*, corporate executives and professional advertisers use the term *segment*. No matter the jargon, what can be seen more clearly after Deleuze is how the threat to identity has swung halfway around: it is no longer that we may be dissolved into some unity larger than ourselves, it is that we may be shattered into shards of diverse interests that, collected, no longer resemble the person who is, presumably, their reference (Hammond 2016: 452–467). That is, our desires and purchases are no longer aspects of a coherent self so much as disconnected spurts of purchasing corresponding with market segments. The fifty-year-old male wants a Land Rover, and the father of school-age children wants a spring break vacation, but there is no necessary connection between those wants even when they correspond with a single conventional identity.

Moving on to the subject of occupations, Deleuze originally foresaw the labours of data analysis and consumer prediction as surging along with the rise of control society and the economic transition from producing to selling. Today, the kind of selling that revolves around predictive analytics seeks *trajectories*: consumers are conceived as always on their way somewhere, and the job is to determine their aim (Ghose 2017: 53). So, a vacationer who scuba dives off the Florida coast is targeted by online ads for a trip to Cozumel in Mexico. After the Cozumel dive, he is lured by a discounted flight to Australia's Great Barrier Reef. Then, while waiting in the Sydney airport for the flight back home, an intriguing article about cave diving crosses his Facebook feed. The advertising of trajectories converts experiences into the desire for another, further down the continuous line. It also enables Deleuze's joy of marketing, which is selling from abundance: you want to go on the next trip not because the last one was less than satisfying, but because it was so enjoyable.

The saga of the iPhone also provides an apt example. Each year's model is not only about what the technology offers now, but also how this new device allows users to anticipate – and advertisers to promote – the *next* version. The result is buyers lining up in front of Apple stores to purchase the new iPhone, but already excitedly speculating about what might be included in the subsequent model (Gianatasio 2012). Services and products are no longer destinations, they're stations, points of departure as much as arrival, and the consequent marketing task is to ensure that experiences aren't ends in themselves but ways of advancing towards the next one (Weilbacher 2003: 230–234).

This marketing logic of trajectories is exceedingly well suited to the control society for three reasons. First, control's basic strategies are activated. When it comes to mapping flows of individuals, recognising patterns, and leading through incentives, today's data-driven advertisers – supported by a vast and growing infrastructure of information collection, processing, and distribution enterprises – are practical experts. Second, the method of social organisation auto-intensifies. If control on the economic level is based on the advertising logic of trajectories, then implementation through personal information gathering and predictive analytics facilitates still more implementation. As marketers grow more adept at bending artificial intelligence and big data into the project of motivating consumers, profits mount, and the increasing resources pour back into the data gathering and processing, and then into deploying still more finely tuned incentives (Kietzmann et al. 2018). Third, those incentives verge on the irresistible because *their coercive force is the consumer's own resolved identity*. Deleuze's joys of marketing tend towards euphoria.

## Programme

'What counts', Deleuze wrote near his essay's end, 'is that we are at the beginning of something'. (Deleuze 1992: 7). That is why, he concluded, it falls to 'the young people to discover what they are being made to serve' as the mechanisms of control society extend.

### *Thirty-year Update*

Deleuze's young people are today's forty to sixty set, and for us much remains to be discovered of the programme dedicated to revealing the risks and opportunities distinguishing lives under control, but at least three clarifications have emerged since the original essay. Each provides guidance for living in the control society as it exists today.

Clarification one involves the role of labour unions which, for Deleuze, posed 'one of the most important questions'. (Deleuze 1992:7). He asked, 'Would they be able to lead struggles against oppression in the control society?' No. Reduced to an anachronism by the rush of today's freelancer economy in the United States and (increasingly) Europe, it almost seems bizarre that only three decades ago it was natural to privilege the social role of the collectives, and to envision unions as potentially radiating force beyond economic activities. Only a few lines further on, Deleuze's readers encounter his glancing reference to the 'rough outlines' of the coming threat named 'the *joys of marketing*'. (Deleuze 1992: 7). Again, here, there is a stark difference: today it is not banal to observe that people watch the Super Bowl because they want to see the ads. In other words, it is no longer union collectives but marketing stars – with their unique joys powered by algorithms, fuelled by data, and predicting the trajectories of resolved identities – who most influentially surge from the economic world into social and cultural reality. So, if we are considering how to live in the control society, the bearings initially sought from organised labour switch towards a requirement for awareness of, and familiarity with, big data marketing strategies.

The second clarification is the adjustment of central examples. Thirty years ago, the distinction between discipline and control adhered well to the difference between the cement walls of a prison and the electronic collar of house arrest. The example had the further virtue of evoking Foucault's notable discussion of Bentham's Panopticon (Kelly 2015: 148–162). Today, however, a more relevant social and cultural distinction today is brick and mortar stores against – and being displaced by – online retailing.

The coercive power of brick and mortar stores culminated as *shopper marketing*, a disciplinary strategy that plateaued in the late 1990s (Ståhlberg and Maila 2012). Based on the manipulation of space inside a physical store, the idea was that customers could be organised by corridors, racks, and displays to make certain kinds of purchases. The most obvious embodiment was the narrow lane of the check-out line: with nothing else to do while standing and waiting, the magazines helped pass the time (maybe you will read half the article, and then buy the issue to finish), and candies were abundant (you just crossed an item off your to-do list, grab your reward). All this is about organising consumers and their purchasing with space and barriers. Contrastingly, online merchants work within a logic of control to track and target consumers in an unbounded purchasing environment. Stores are everywhere (as long as you have your phone and are internet connected), and you are constantly in the checkout line (two clicks buys any product appearing on the screen). The coercive potential is unlimited: you *always* may be deserving, there is *always* something that may be needed. The art of targeting in online marketing and retailing is the conjuring of that merit and desire, and when it is functioning, every moment of every day becomes a retailing opportunity. We are all together in the interminable market, though each of us has our own trajectory, and next purchase to make.

The final update for living in a control society involves privacy, defined as individuals having the power to regulate access to their own personal information (Westin 1968: 3). The subject barely registers in a disciplinary society where the physical apparatus – the prison wall, the office door, the room of a compartmentalised home – minimises the subjects' personal uniqueness. Disciplinary overseers do not need access to private data, they do not need to know the individualising traits, habits, and tastes of their various wards. The only requirement is a space where eating happens, and the ability to march subjects into the cafeteria. Contrastingly, if organising happens as control, and if the central mechanisms are the tracking technologies and luring encouragements of predictive analytics, then those exercising control need to learn particular hungers, capabilities, vulnerabilities, aspirations, fears, and hopes. The value of *knowing* about people surges as incentives replace walls, because it is personal data that perfects algorithmically deployed incentives (Palmås 2011). So, three decades after Deleuze, a subject he did not mention – personal information data sets – moves to the centre of struggles in control society.

Already in 1999, Scott McNealy, head of Sun Microsystems, announced, 'You have zero privacy anyway, get over it' (Sprenger 1999). Today, the resignation seems even more inevitable. Since nearly everything we do connects with digital applications, almost every part of ourselves is susceptible to data conversion, to packaging for sale, and to uncontained dissemination on the information markets (Miley 2018). The extent of the exposure is suggested by the notorious case of Target stores compiling consumer purchasing patterns with sufficient expertise to realise a young woman's pregnancy so they could send maternity advertisements to her home address. A scandal ensued when the news surprised her family. In fact, since the conclusion was gleaned from purchases that had no direct relation to motherhood (a particular mix of products including several lotions, not a pregnancy test) it is possible that Target knew about the next generation even before the mother herself (Trout 2017).

So, what is the privacy struggle today? For humans vulnerable to data collection, at least three fronts exist. One involves blocking access to our personal information through technical and regulatory measures. Web browsers can be set to private mode, digital engineers can adopt privacy by design initiatives, legislators can promulgate measures including the General Data Protection Regulation. A distinct kind of effort seeks to preserve what remains of individual privacy not by limiting the information supply, but by oversupplying. This strategy of obfuscation mixes an abundance of false data in with the truth, so as to render the legitimate information unlocalisable even while visible (Brunton and Nissenbaum 2016). The third front is the most extreme and speculative, and is explored in this essay. It is that privacy can no longer be saved, so the struggle is about a direction after the loss. Here, the vital question for the programme of response to control is: *What is to be done now that privacy is gone?*

### Practical Conclusion on Privacy Lost, and Control

Envisioning life after privacy has always been possible, but it has never been inevitable as it is now, under the pressure of the control society. Ultimately, control's most significant historical contribution may be what it provokes: the particular coercion of big data and predictive analytics operating on the economic level encourages or even *produces* an exploration of post-privacy human experience.

Theoretically, there are two limiting extremes to the exploring, though in the practical terms of actual human lives their viability remains an open question. They are: transparency and discontinuity.

Transparency means indiscriminately displaying everything there is to know about ourselves. The possibility is shocking, but also enticing. If we can get satisfying experiences by proactively providing all-access passes to the personal details of our lives instead of waiting for the data to be squeezed out of our daily routines and digital exhaust, then shouldn't we rush to expose everything to the matchmakers at Tinder, and to the headhunters at LinkedIn? Stronger loves on demand, and better jobs industriously filtered for our inbox is a good start. Then there is more to be had as the technologies of human-tagging (Voas and Kshetri 2017) and überveillance (Michael et al. 2008) advance. How much will parents pay – in terms of intimate personal information – to find a vacation that truly
brings their family together? How much unfiltered biological information
will the youthful and the elderly transmit to healthcare enterprises –
perhaps via an implanted microchip – in return for a guaranteed alert one
year before a cancer becomes inoperable, or one hour before a heart attack
(Raghupathi and Raghupathi 2014)?

If the answer is everything – total transparency – than we're touching
the marketers' dream where consumer habits, urges, conditions, and
desires become so visible that they can be answered *even before* they are
realised: satisfaction precedes wanting, and escape from contentment
becomes impossible. Convenience becomes euphoric.

Personalised service rising to meet the commitment to transparency
also means surging control: when deployed incentives are calibrated by
perfect knowledge of their subjects, it is difficult to see opportunities for
deviation from the algorithmically oriented trajectories (Palmås 2015) of
completely resolved identities. The incentives' allure will get underneath
and overwhelm individual choice and autonomy by enticing consumers to
be the person *their own* complete data set demands. The control society
raised to the ideal means every subject is forced to be exactly who they are.
There is, that means, a divergence between freedom and authenticity in the
perfected control society: as the former evaporates, the latter crystallises.

The second extreme response to privacy's absence is not about
revealing, but *transforming* one's own personal information. The idea is to
render all the privacy-violating data that has been accumulated
inapplicable. If particular consumers and social media users manage to
abruptly shift their tastes, aversions, and urges, then the personal
information previously gathered to profile them, along with the
corresponding targeting algorithms and deployed incentives will twist
nonsensically. Control will recede – at least momentarily – because no
matter how much identifying data marketers have catalogued, their
predictive analytics will misfire when aimed for someone who has already
become someone else.

The ideal existence becomes discontinuous, and *becoming-other* in Deleuze's language (Semetsky 2011) drives a strategy that starts by severing the past as an effect, not a cause, of differentiation. Accumulated personal information is not escaped so much as discarded along the way forward just as a rhizome plant stem breaks away from the longer root (Deleuze and Guattari 1987). Ontologically, this regeneration corresponds with the privileging of difference over identity at the production of being itself (Deleuze 1983), but what's more humanly pressing is this question: What experiences create immunity from past data – for a consumer, for a user, for a person – in the time of surveillance capitalism (Zuboff 2019)?

Personal changes sufficiently powerful to crack big data identity profiles are rare but familiar. Ordinarily stable people readjust who they are seismically when they go away to college, marry, have children. Think, for example, of the kinds of things people would be predicted to do – and could be incentivised to want – around midnight at each of those stages. On a different level, there are biological examples of personal transformation including gender reassignment. On the cultural level, the travel section of any bookstore includes stretches of titles dedicated to experiences resembling those of Paul Bowles' *Sheltering Sky*, episodes where people go abroad and recreate themselves as attuned to customs and satisfactions that are as foreign as the new language they are learning.

Other opportunities for immunity from accrued personal information are enabled by the very platforms and technologies that threaten to build inescapable identity data banks. Take LinkedIn displaying job openings that are curated for our digitally modelled self. To the extent our future career is channelled by predictive analytics churning our own past datapoints (Staddon 2009), the platform is an ingenious machine of control: not only are users coerced without prohibitions, the directives arrive as professional *opportunities*. At the same time, though, and with a little search roguery, job counter-opportunities can be discovered in cities and professions that our past information would have otherwise excluded. These may be places we have never visited, governed by expectations and aspirations outside the profile of our resolved identity: the Silicon Valley entrepreneur accepts a corporate position in the Swedish welfare state, the feminist purchases a headscarf and a one way ticket to an English-teaching job in Saudi Arabia. Regardless, life-jarring opportunities are out there. LinkedIn makes then accessible, and for those who transmit a resonant appeal to the recruiter who's willing to take a chance, they can be gone the next day.

Romance platforms provide a similar dynamic: with some counterfeit personal information and creative swiping, the algorithms designed to locate data-verified compatible matches can be perverted to generate encounters entirely detached from the familiar rules of attraction. Where that may lead is difficult to foresee, but what is critical is the potential. The platforms gathering information and spreading coercion also enable regenerated data about tastes, values, vulnerabilities, and desires that crack identity profiles, that crash algorithms, and that lose control.

Thirty years from now the cultural horizon may no longer be dominated by big data and predictive analytics. Today, though, Deleuze's original imperative to determine what we are being made to serve as control advances leads to struggles over our personally defining information. The struggles will lead somewhere between the transparency belonging to those who embrace the joys of marketing, and the transience of those cutting away from their own identity.

**End**

## Acknowledgements

Anindya Ghose, Riehl Professor of Business at New York University, contributed important insights to this article in the area of big data and predictive analytics.

Vishal Lala, Lubin School of Business at Pace University, contributed important insights to this article in the area of contemporary data analytics, and marketing.

Early and partial versions of this article were presented at CEPE (Computer Ethics –Philosophical Enquiry) 2019: Norfolk, Virginia, and The Society for Business Ethics 2019: Boston, Massachusetts.

## Author Biography

James Brusseau (PhD, Philosophy) is author of books, articles, and digital media on the history of philosophy and ethics. He has taught in Europe, Mexico, and currently at Pace University near his home in New York City. As Director of the AI Ethics Site, a research institute currently incubating at Pace University, he explores the human experience of artificial intelligence. E-mail: jbrusseau@pace.edu